\documentclass[conference]{IEEEtran}
\ifCLASSINFOpdf
\else
\fi
\usepackage[table]{xcolor}
\usepackage{wrapfig}

\usepackage{tikz}
\usepackage{graphicx}
\usepackage{comment}
\usepackage{floatrow}
\usepackage{amssymb}
\usepackage{amsmath} 
\usepackage{color} 
\usepackage[all]{nowidow}
\usepackage[font=small,skip=0pt]{caption} 

\usepackage{epsfig,endnotes}
\usepackage{wrapfig}
\usepackage{tikz}
\usepackage{fontawesome}
\usepackage{amssymb}
\usepackage{multirow}
\usepackage{mathtools}
\newcommand\system{\texttt{\sc EnTrust}}

\newcommand\circled[1]{\tikz[baseline=(char.base)]{
            \node[shape=circle,draw,color=white,fill=black,inner sep=1pt, minimum size=11pt](char){#1};}}

\newcommand\squared[1]{\tikz[baseline=(char.base)]{
            \node[shape=rectangle,draw,color=white,fill=black,inner sep=0pt, minimum size=9pt](char){#1};}}

\newcommand\marked[1]{\tikz[baseline=(char.base)]{
            \node[shape=rectangle,draw,color=black,fill=lightgray,inner sep=0pt, minimum size=9pt](char){#1};}}

\newcommand\bluemarkA[1]{\tikz[baseline=(char.base)]{
            \node[color=black,fill=blue!10,inner sep=0pt, minimum size=9pt](char){#1};}}

\newcommand\bluemarkB[1]{\tikz[baseline=(char.base)]{
            \node[color=black,fill=blue!40,inner sep=0pt, minimum size=9pt](char){#1};}}

\newcommand\yellowmarkA[1]{\tikz[baseline=(char.base)]{
            \node[color=black,fill=yellow!20,inner sep=0pt, minimum size=9pt](char){#1};}}

\newcommand\yellowmarkB[1]{\tikz[baseline=(char.base)]{
            \node[color=black,fill=yellow!60,inner sep=0pt, minimum size=9pt](char){#1};}}

\begin{document}
%
\title{Regulating Access to System Sensors\\in Cooperating Programs}

\author{\IEEEauthorblockN{Giuseppe Petracca}
\IEEEauthorblockA{Penn State University, US\\
gxp18@cse.psu.edu}
\and
\IEEEauthorblockN{Jens Grossklags}
\IEEEauthorblockA{Technical University of Munich, DE\\
jens.grossklags@in.tum.de}
\and
\IEEEauthorblockN{Patrick McDaniel}
\IEEEauthorblockA{Penn State University, US\\
mcdaniel@cse.psu.edu}
\and
\IEEEauthorblockN{Trent Jaeger}
\IEEEauthorblockA{Penn State University, US\\
tjaeger@cse.psu.edu}}

\maketitle

\begin{abstract}
Modern operating systems such as Android, iOS, Windows Phone, and Chrome OS support a cooperating program abstraction. Instead of placing all functionality into a single program, programs cooperate to complete tasks requested by users.
However, untrusted programs may exploit interactions with other programs to obtain unauthorized access to system sensors either directly or through privileged services.
Researchers have proposed that programs should only be authorized to
access system sensors on a user-approved input event, but these methods do
not account for possible delegation done by the program receiving the user input event. Furthermore, proposed delegation methods do not enable users to control the use of their input events accurately.
In this paper, we propose \system, a system that enables users to
authorize sensor operations that follow their input events, even if
the sensor operation is performed by a program different from the program receiving the input event.  \system\ tracks user input as well as delegation events and restricts the execution of such events to compute unambiguous delegation paths to enable accurate and
reusable authorization of sensor operations.
To demonstrate this approach, we implement the \system\ authorization system for Android. We find, via a laboratory user study, that attacks can be prevented at a much
higher rate (54-64\% improvement); and via a field user study, that \system\ requires no more than three additional authorizations per program with respect to the first-use approach, while incurring modest performance (\textless 1\%) and memory overheads (5.5 KB per
program).
\end{abstract}

\section{Introduction}

Modern operating systems support a programming abstraction that enables programs to cooperate to perform user commands.  Indeed, an emergent property of modern operating systems is that applications are relatively simple, provide a specific functionality, and often rely on the cooperation with other programs to perform a larger task. For instance, modern operating systems now ship with powerful
voice-controlled personal assistants that are entitled to interface with other programs, reaching for a new horizon in human-computer interaction.  

Unfortunately, these assistants are valuable targets for adversaries. Indeed, real-world scenarios have demonstrated how voice-controlled personal assistants are vulnerable to information leaks~\cite{kastrenakes_2017, bhattacharya_2017}. Likewise, researchers have shown how simple voice commands, either audible or inaudible, can be used to control personal assistants and mislead them  in various ways~\cite{zhang2017dolphinattack, diao2014your, alepis2017monkey, petracca2015audroid}.  Just to mention one of the most recent cases, a ride-sharing app took advantage of an entitlement provided by a service part of Apple iOS, as recently reported by Gizmodo \cite{conger_2017}. Whenever users asked their voice assistants ``Siri, I need a ride'', the assistant launched the ride-sharing app. Leveraging the entitlement provided by the service, the app requested  recording of users' screens, even while running in the background. The ride-sharing app abused this service to spy on users, while they were taking a ride.  
  
However, voice assistants are just one type of application that may be exploited by such types of attacks.  Several real-world incidents have been reported to the public, where malicious apps, able to track and spy on users by stealthily opening smartphones' cameras, microphones, and GPS receivers, made their way through official stores such as the Apple App Store and the Google Play Store, after bypassing strict pre-publishing security analyses~\cite{apple1, apple2, android1, googleblog2017}.  In such attacks, the user is victimized when adversaries obtain unauthorized access to the user's data by tricking a voice assistant to use an untrusted program or by tricking the user into interacting with malicious programs.

\begin{figure*}[ht!]
\centering
\includegraphics[width=175mm]{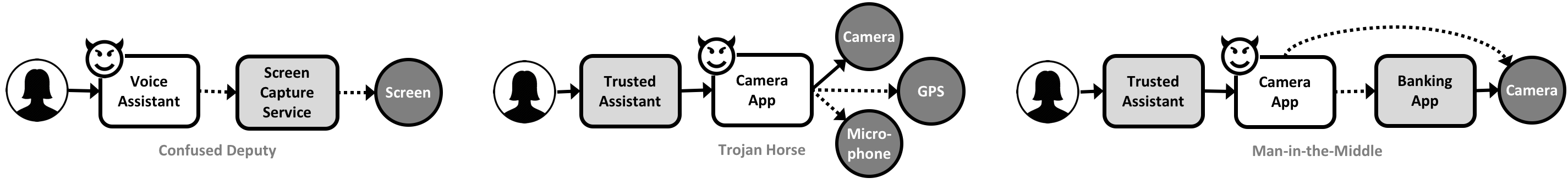}
\caption{Possible attack vectors when considering diverse programs interacting in a cooperating model.}
\label{fig:vectors}
\end{figure*}

Researchers have proposed methods that enable users to authorize how programs may access system sensors by binding user input events to sensor operations \cite{udac, udac2, Petracca2016AwareCA, petracca2017aware, overhaul}, but
these methods do not account for programs enlisting other programs to perform sensor operations.  Thus, programs that receive user input events may exploit or be exploited by cooperating programs to perform unauthorized sensor operations. Other studies focused on mechanisms to regulate Inter-Process Communication (IPC), which can track and control delegations~\cite{chin2011analyzing, li2015iccta, felt2011permission, octeau2015composite, octeau2016combining, aafer2015hare}; however, these mechanism are 
unable to account for the user's intention for the input events that they generate.  As a result, malicious programs may trick users to perform unauthorized sensor operations.

In this work, we propose the \system\ authorization system to prevent
malicious programs from exploiting cooperating programs to obtain unauthorized access to system sensors.  The main challenge is to connect user input events made to one program to sensor operation requests made by another program unambiguously.  To do this, \system\ tracks user input events and inter-process communications to construct a {\em delegation graph} that \system\ uses for authorizing each sensor operation request.  On a sensor operation request, \system\ uses the delegation graph to compute the {\em delegation path} from the program making the operation request back to the program that elicited the original user input event, and presents that path to the user for authorization of these programs.  \system\ leverages the insight that user input events are relatively infrequent, processed much more quickly than users can generate distinct events, and are higher priority than other processing, to delay events that may result in multiple feasible delegation paths.  \system\ then reuses authorizations for delegation paths to reduce user authorization effort.  Thus, \system\ enables users to ensure that their expectations are met, even if a group of programs collaborate to process a user command by using sensor data~\cite{jung2012short,lin2012expectation}.  

We implement and evaluate a prototype of the \system\ authorization system on a
recent version of the Android operating system, and find that
\system\ prevents three attack vectors possible in systems supporting cooperating programs, requires little additional user
effort, and has low overhead in performance and memory consumption.
In a laboratory user study involving 44 human subjects,
\system\ reduces users' failure to detect attacks by 54-64\% when compared to
first-use authorization.  In a field user study involving 9 human subjects,
we found that users required no more than three additional manual authorizations per
program.  Lastly, we measured the overhead imposed by \system\ via
benchmarks and found that programs operate effectively under
\system, while incurring a negligible performance overhead 
 and a memory footprint of only 5.5 kilobytes, on
average, per program based on macro-benchmarks. 

In summary, in this paper we make the following contributions:

\begin{itemize}
    \item We identify three types of attack vectors exploitable by malicious programs attempting to gain unauthorized access to system sensors by leveraging cooperation with other programs. 
    \item We propose \system, an authorization system that computes unambiguous \textit{delegation paths}, from programs eliciting user input events to programs requesting sensor operations. As such, \system\ 
      enables users to authorize sensor operations resulting from those
      delegations and systems to reuse those authorizations for
      repeated requests.
    \item We implement the \system\ prototype and test its
      effectiveness with a laboratory-based user study, the users'
      authorization effort with a field-based user study, and
      performance and memory overhead via benchmarks.
\end{itemize}

\section{Problem Statement} 
\label{sec:problem}

Users initiate actions targeting system sensors without having control over \textit{which programs} are going to service their requests. After users authorize programs to perform sensor operations via first-use authorization, users have no control over \textit{when} programs are going to use permissions to perform sensor operations in the future.  Unfortunately, this lack of control can lead to privacy violations when such operations target system sensors, such as cameras, microphones, GPS receivers, or the content displayed on users' screens. 
For instance, in first-use authorization, users have choices about what programs they install and what permission they grant; however, after first-use, programs are free to do \textit{whatever} they want with the acquired permissions and use them \textit{whenever} they want. A step forward was made with the introduction of runtime permissions, which allow users to revoke programs' permissions after installation. Unfortunately, users have still no visibility of \textit{which programs} apply their permissions following a user input event and \textit{when} programs use granted permissions. Thus, realizing when programs abuse granted permissions, to decide whether to revoke them, remains a challenge for users.

Due to the lack of control over what happens between user input events and the resulting operation requests, malicious programs may leverage the following three attack vectors. For consistency, we will present the attack scenarios in terms of voice assistants receiving user input events via voice commands; however, similar attack scenarios are possible for input events received by programs via graphical user interface widgets rendered on the screen.

\textbf{Confused Deputy} --- A malicious program may leverage a user input event as an opportunity to confuse a more privileged program into performing a dangerous action, as in the ride-sharing case mentioned above. However, other scenarios are possible. For instance, a third-party voice assistant app\footnote{\scriptsize There exist over 250 voice assistant apps on Google Play developed by unknown app developers.} may be maliciously calling the screen capture service every time that a user requests to create a voice note. The malicious app may therefore succeed in tricking the screen capture service to capture and leak sensitive information (e.g., a credit card number written down in a note). All the user can see is the new note created by the notes app. This is an instance of a \textit{confused deputy attack} in a cooperating scenario, as depicted on the left side of Figure~\ref{fig:vectors}.

\textbf{Trojan Horse} --- A program trusted by the user may delegate the received input to another program able to perform the requested action. For instance, a trusted voice assistant may activate a camera app to serve the user request to take a selfie. However, the camera app may be a Trojan horse app that takes a picture, but also records a short audio via the microphone, and the user location via GPS (e.g., a spy app\footnote{\scriptsize One of the many spy/surveillance mobile apps available for purchase online (e.g., \texttt{flexispy.com}).} installed by a jealous boyfriend stalking on his girlfriend). All the user can see is the picture being taken by the camera app. This is an instance of a \textit{Trojan horse attack} in a cooperating scenario, as shown in the middle of Figure~\ref{fig:vectors}.

\textbf{Man-In-The-Middle} --- A program trusted by the user may be leveraged to spoof the user into initiating an interaction with an unintended program. For instance, a legitimate banking app may leverage the voice interaction intent mechanism\footnote{\scriptsize{Add Voice Interactions to Your App (https://codelabs.developers.google.com/)}} to allow clients to launch the banking app by asking their voice assistant to ``deposit check.'' However, a malicious program may define a similar voice interaction, such as ``deposit \textit{bank} check,'' with the voice assistant.  Therefore, what would happen is that whenever the user instantiates the ``deposit \textit{bank} check'' voice command, although the user expects the legitimate banking app to be activated, the malicious app is activated instead. The malicious app opens the camera, captures a frame with the check, and finally sends a spoofed intent to launch the legitimate banking app, all while running in the background. All the user can see is the trusted banking app opening a camera preview to take a picture of the check. This is an instance of a \textit{man-in-the-middle attack} in a cooperating scenario, as shown on the right side of Figure~\ref{fig:vectors}. This is only one of several ways a user can be spoofed. Indeed, researchers have reported several ways programs can steal and spoof intents \cite{aafer2015hare, chin2011analyzing, huang2012clickjacking, luo2012touchjacking, touchjacking}. 

Mechanisms solely based on permissions \cite{felt2011permission, Felt:2012, felt, wijesekera2017feasibility} to regulate access to system sensors are not sufficient to prevent these attack vectors. First, a malicious program may leverage trusted programs as confused deputy, thus exploiting the deputies' permissions. Second, the malicious program may abuse already granted permissions. For instance, it is reasonable to grant a camera app permission to access the device camera, the microphone, and geo-location information. However, it is less acceptable for this camera app to access the geo-location information and the microphone when the user only intended to take a selfie.

Prior work binds user inputs to programs' access to system sensors \cite{udac, udac2, overhaul, Petracca2016AwareCA, petracca2017aware} to prevent programs from stealthily accessing sensors or from performing operations different from those intended by the users. However, they are unable to track delegations that may follow user input events. Furthermore, prior work that studied mechanisms to track Inter-Process Communication (IPC) \cite{chin2011analyzing, li2015iccta, octeau2015composite, octeau2016combining, aafer2015hare} may enforce policies to prevent hijacking or stealing of intents. However, policies may be incomplete (e.g., unable to account for unexpected interactions) and unable to account for the user's intention in diverse scenarios. Also, mechanisms focused on preventing permission re-delegation \cite{felt2011permission}, whenever a program with more permissions services other programs' requests, are too restrictive and inadequate for systems where programs are meant to cooperate to serve requests from users. 

Lastly, mechanisms based on taint analysis \cite{Enck2010, arzt2014flowdroid, tang2012cleanos, li2015iccta} or Decentralized Information Flow Control (DIFC) \cite{nadkarni2013preventing, nadkarni2016practical} are able to, respectively, track and control how sensitive data is used by or shared between programs; however, such mechanisms solve the orthogonal problem of controlling sensitive data leakage. They do not solve the problem of establishing whether access to system sensors leads to sensitive data that users think would cause privacy violations. 

\textbf{Security Guarantee}: To block the attack vectors described above, we must design a defense mechanism able to provide the following security guarantee: 

\textit{Any sensor operation must be initiated by a program receiving a user input event, directly or indirectly through IPC, and the user must authorize the entire sequence of involved programs, from the one that received the user input event to the one requesting the sensor operation.}

\section{Security Model}
\label{sec:model}

\indent\textbf{Trust Model} -- We assume that the system (e.g., Linux kernel, operating system, system services, and device drivers) is booted securely, runs approved code from device vendors, and is free of malice. We assume that user-level programs (e.g., applications) are isolated from each other via the sandboxing mechanism using separated processes~\cite{sandboxing1, sandboxing2}. We assume that, by default, user-level programs have no direct access to system sensors due to the use of a Mandatory Access Control (MAC) policy \cite{smalley2001implementing, smalley2013security} enforced from boot time. We assume the use of {\em trusted paths}, protected by MAC, allowing users to receive unforgeable communications from the system, and providing unforgeable user input events to the system. These assumptions are in line with existing research proposing trusted paths and trusted user interfaces for browsers~\cite{trustedpathbrowser}, X window systems~\cite{zhou2012building, shapiro2004design}, and mobile operating systems~\cite{li2014building}. 
 
\textbf{Threat Model} -- We assume that users may install programs from unknown sources (potentially malicious), then grant such programs access to system sensors at first-use. Despite the default isolation via sandboxing, programs may communicate via message passing mechanisms (i.e., intents or broadcast messages). Thus, user-level programs (e.g., applications) may leverage such communication to exploit any of the three attack vectors described in Section~\ref{sec:problem} and depicted in Figure~\ref{fig:vectors}.
Our objective is to regulate how cooperating programs access system sensors. How programs manage and share the data collected from system sensors is outside the scope of our research. Our objective is not to prevent data leakage from or denial of service by compromised or colluding programs. Researchers have already examined solutions to prevent data leakage based on taint analysis \cite{Enck2010, arzt2014flowdroid, tang2012cleanos, li2015iccta} and decentralized information flow control \cite{nadkarni2013preventing, nadkarni2016practical}.

\section{\system\ Approach Overview}
\label{sect:overview}

\begin{figure}[t]
\centering
\includegraphics[width=80mm]{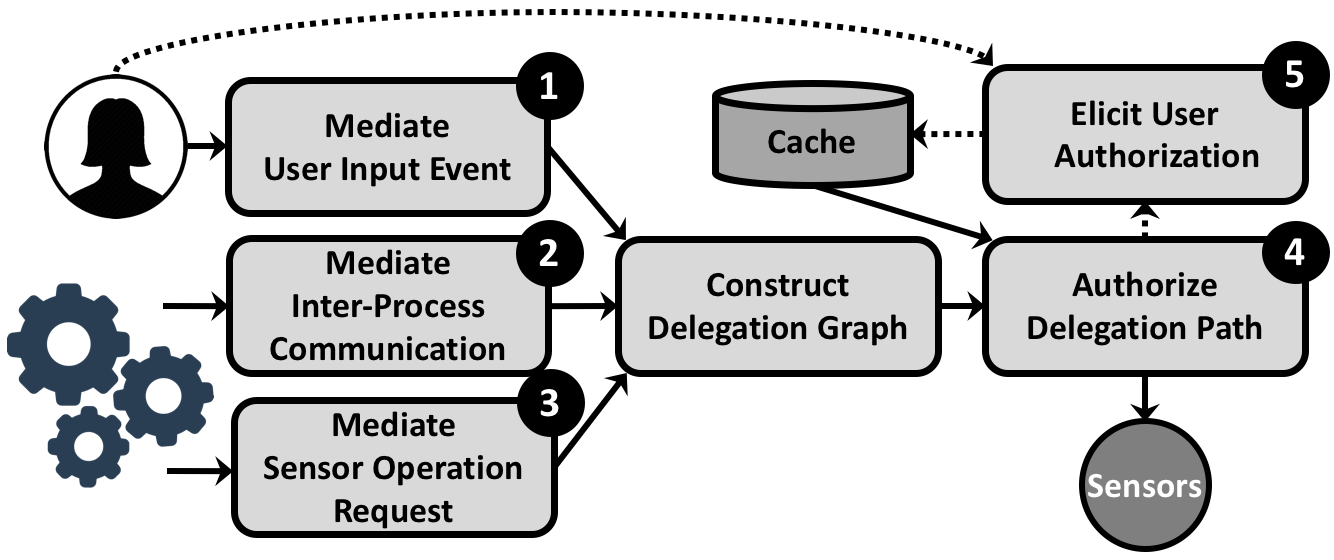} 
\caption{{\bf \system\ authorization method}: user input events, handoff events, and sensor operations are linked via delegation graphs to compute unambiguous delegation paths for user authorization of sensor operations.}
\label{fig:strawman}
\end{figure}

\system\ restricts when a program may perform a sensor operation by requiring each sensor operation to be unambiguously associated with a user input event and an authorization for that sensor operation, even if the sensor operation is performed by a program different from the one receiving the user input event.  

Figure~\ref{fig:strawman} provides an overview of the \system\ authorization system, which consists of five steps.  In the first three steps, \system\ mediates and records user input events, inter-process communication events (handoff events), and sensor operation requests, respectively, to construct a {\em delegation graph}. All of these steps are performed in a manner to ensure that user input events can be associated unambiguously with sensor operation requests.  In the fourth step, \system\ uses the constructed delegation graph to compute an unambiguous {\em delegation path} from the input event to the sensor operation request.
Unless the authorization cache contains a user authorization for this delegation path already, the fifth step elicits an authorization for the delegation path from the user, and caching of user authorizations for later use for the same delegation path.

\begin{figure}[h!]
\centering
\includegraphics[width=52mm]{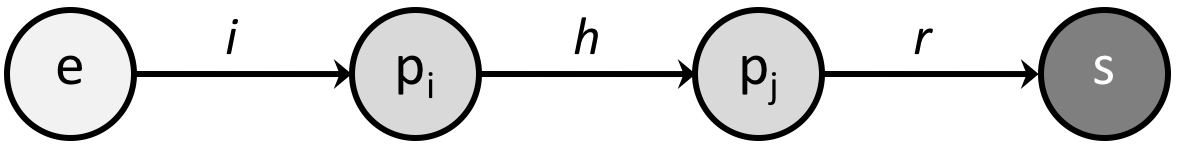}
\caption{Delegation graphs connect input events with operation requests for sensors via handoff events.}
\label{fig:delegation-path}
\end{figure}

\system\ connects sensor operation requests to user input events via handoffs by constructing a delegation graph to record such operations, as shown in Figure~\ref{fig:delegation-path}.  A delegation graph is a graph, $G=(V,E)$, where the vertices, $v \in V$, represent user input events, programs, or system sensors and the edges $(u,v) \in E$ represent the flow of user input events to programs and system sensors.  Figure~\ref{fig:delegation-path} shows a simple flow, whereby a user input event $i$ is delivered to a program $p_i$ that performs a handoff event $h$ to a program $p_j$, which performs an operation request $r$ for sensor $s$. Thus, there are three types of edges: user input event to program (user input delivery), program to program (handoff), and program to sensor operation request (request delivery).  

Upon a sensor operation request, \system\ elicits an explicit user authorization for the path in the delegation graph associated with the user input event that led to the operation request, called the {\em delegation path}. \system\ finds the exact delegation path by traversing the delegation graph backwards from the operation request to the user input event that led to the request. The identified delegation path can be expressed in simple natural language, similar to first-use authorizations, and we assess how the user utilizes delegation paths to produce authorizations in a user study in Section~\ref{sec:user_study}. 

Critical to computing delegation paths is the ability for \system\ to find an unambiguous reverse path from the sensor operation request back to a user input event.  There are two main challenges.  First, a program may receive multiple user input events, so it may be ambiguous as to which user input event led to a handoff and/or sensor operation request.  Second, a program may receive multiple handoffs from other programs and it may be ambiguous which input event led to a sensor operation request.  In particular, a delegation path is said to be \textit{unambiguous}, if and only if given an operation request $r$ for a sensor $s$ in program $p_j$, it is possible to identify a single user input event $i$ for program $p_j$ that preceded it or it is possible to identify a single path $p_i \rightarrow p_j$ in the delegation graph and $p_i$ receives a single input input $i$.  Without requiring changes to system APIs and program code to use them, resolving such ambiguities at the system level is not practical, unless some restrictions are enforced on how programs operate.

Our insight is based on the recognition that user input events are relatively infrequent, processed much more quickly than users can generate them, and higher priority than other processing. Thus, we design \system\ to enforce restrictions to prevent the creation of ambiguities in delegation graphs.  First, we limit programs to receive only one input event at a time.  That is, from the time that a user input event is received by a program until the completion of processing resulting from that input, no other distinct user input events may be delivered to that program.\footnote{\scriptsize The same type of user input event may be delivered, but it must result in the same delegation path.}  Second, to determine when processing must complete, we bound processing time from the initial delivery of a user input event based on the amount of time needed for users to generate the next user input event.  We observe and measure, as described in Section~\ref{sec:no_ambiguous}, that distinct user input events can only be created at a much slower rate than sensor operation processing.  Third, to resolve ambiguities due to handoffs, we restrict receivers of a handoff, originating from a user input event, to only one active handoff event.  While this may sound like a big lock over programs, we mitigate this impact by prioritizing handoffs from user input events over others to "get them out of the way."  Fortunately, this approach complies with typical scheduling expectations that I/O events are prioritized over CPU events.

\section{\system\ Authorization Design}
\label{sect:design}

In this section, we describe a method for authorizing delegated sensor
operations.  This method solves three problems.  First,
\system\ tracks user input events, handoff events, and sensor operation requests to build delegation graphs (Section~\ref{sec:input-and-handoff}).
Second, \system\ computes delegation paths
between any user input event and any sensor operation request unambiguously (Section~\ref{sec:no_ambiguous}).  Third, \system\ enables users to authorize constructed delegation paths while approximating the first-use authorization in terms of user effort (Section~\ref{sec:authorization}).

\subsection{Building Delegation Graphs}
\label{sec:input-and-handoff}

We start by defining the basic methods used by \system\
to mediate {\em user input events}, {\em handoff events}, and {\em sensor operation requests} necessary to construct \textit{delegation graphs} rooted at input events.  

First, for each user input event for a program $p_i$, \system\ creates an \textit{input event} tuple $i=(w,p_i,t_i)$, where $w$ is the widget used to
produce the event (including the user interface context, based on prior work~\cite{petracca2017aware});
$p_i$ is the program displaying its graphical user interface on the screen, thus, receiving the input event; and $t_i$ is the time of the user input event (step~\circled{\scriptsize{\textbf{1}}} in
Figure~\ref{fig:strawman}). \system\ is designed to mediated both user input events via GUI widgets as well as voice commands. Voice commands are translated into speech by leveraging the Google Cloud Speech-to-Text service.

Second, after receiving the user input event $i$, program $p_i$ may handoff the event to another program $p_j$. \system\ mediates handoff events by intercepting spawned intents and messages exchanged between programs ~\cite{eugster2003many}. Thus, \system\ models the \textit{handoff event} at time $t_j$ as a tuple $h=(p_{i}, p_{j}, t_j)$,
where: $p_{i}$ is the program delegating the input event, $p_j$ is the program receiving the event, and $t_j$ is the time the event delegation occurred (step~\circled{\scriptsize{\textbf{2}}} in
Figure~\ref{fig:strawman}).

Third, when the program $p_j$ generates a sensor operation request $r$, \system\ models it as a tuple $r=(p_j, o, s, t_k)$, where $p_j$ is the program requesting the sensor operation, $o$ is the type of sensor operation requested, $s$ is the target sensor, and
$t_k$ is the time the sensor operation request occurred (step~\circled{\scriptsize{\textbf{3}}}
in Figure~\ref{fig:strawman}).  

Lastly, \system\ links together all the mediated events to construct a \textit{delegation graph}  $G=(V,E)$, where the $v \in V$ represents an input event, program, or system sensor; whereas $(u,v) \in E$ represents the delivery of user input events, handoffs, and operation requests.  

\subsection{Ensuring Unambiguous Delegation Paths}
\label{sec:no_ambiguous}

Upon mediation of a sensor request $r$, \system\ computes the associated delegation path by tracing backwards from the sensor request to the original user input event $i$. Hence, the operation request $r=(p_j, o, s, t_k)$ above causes a delegation path: ($w,p_i,t_i$) $\rightarrow$ ($p_i,p_j,t_j$) $\rightarrow$ ($p_j,o,s,t_k$) to be reported in step~\circled{\scriptsize{\textbf{4}}} in
Figure~\ref{fig:strawman}.  Delegation paths are then presented to the user for authorization as described in Section~\ref{sec:authorization}.

\begin{figure}[h]
\centering
\includegraphics[width=65mm]{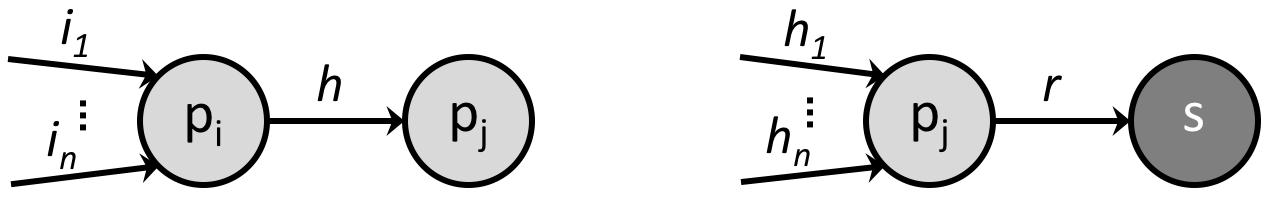} 
\caption{Two scenarios that create ambiguity. Multiple input events or handoff events delivered to the same program.}
\label{fig:ambiguity-cases}
\end{figure}

However, ambiguity is possible in the computation of a delegation path.  We identify two causes of ambiguity.  First, ambiguity is possible if the same program $p_i$ receives multiple input events and performs a handoff, as depicted by the left side of Figure~\ref{fig:ambiguity-cases}. In this case, it is unclear which one of the input events resulted in the handoff.  Second, ambiguity is possible if the same program $p_j$ receives multiple handoff events before performing a sensor operation request, as depicted by the right side of Figure~\ref{fig:ambiguity-cases}.  In this case, it is unclear which handoff is associated with a subsequent sensor operation request.

To prevent these ambiguous cases, we leverage the insights that user input events are relatively infrequent, processed much more quickly than users can generate them, and higher priority than other processing. We leverage these insights to produce a method that ensures sensor operation requests can be connected to user input events unambiguously.  We first describe how to prevent ambiguity for user input events, then we extend the mechanism to handoff events. Also, we leverage scheduling to prevent the proposed mechanism from being used for denial of service attacks against user input event processing.  We evaluate the performance impact of our proposed solution in Section~\ref{sec:performance}.

\textbf{Ambiguity Prevention Mechanism: User Input Events} --  We observe that the time between distinct user input events is much larger than the time needed to produce the corresponding operation request. If every user input event results in an operation request before the user can even produce another distinct user input event, then no ambiguous paths can be generated.  Thus, we propose to set a time limit for each input event, such that the difference between the time at which a user input event is generated $t_i$ and the time for any sensor operation request -- based on that input event -- must be below that limit for the event to be processed.  

Note that repeated user input events (e.g., by holding down a button) are allowed since these repeated user input events generate the same delegation path. Hence, the sequence of sensor operations will be allowed after the delegation path is authorized by the user.  Should the programs produce a different delegation path in the middle of a sequence of operations spawned in this manner, then \system\ would require a new authorization for the new delegation path, as described in Section~\ref{sec:authorization}.

\begin{figure}[h]
\centering
\includegraphics[width=45mm]{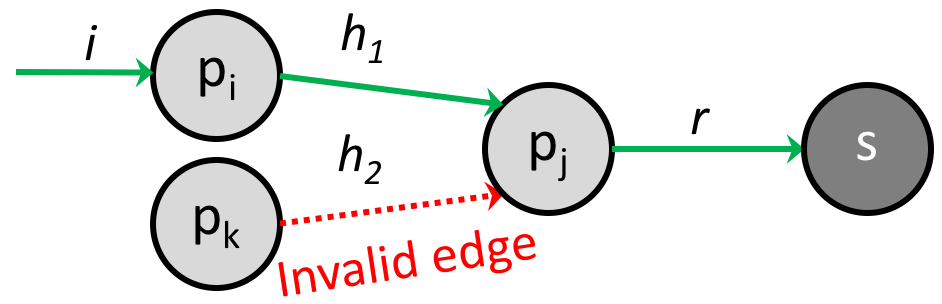} 
\caption{A program $p_k$ attempts leveraging the input event received from program $p_i$ to get program $p_j$ to generate an operation request.}
\label{fig:forbidden}
\end{figure}

\textbf{Ambiguity Prevention Mechanism: Handoff Events} -- Ambiguity prevention for handoff events is more subtle, but builds on the approach used to prevent ambiguity for input events.  Figure~\ref{fig:forbidden} shows the key challenge.  Suppose a malicious program $p_k$ tries to "steal" a user authorization for a program $p_j$ to perform a sensor operation by submitting handoff event that will be processed concurrently to the handoff event from another program $p_i$ that received a input event. Should a sensor operation request occur, \system\ cannot determine whether the sensor request from $p_j$ is to process $h_i$ or $h_k$, so it cannot determine the delegation path unambiguously to authorize the request.  If \system\ knows the mapping between actions associated to handoff events and whether they are linked to sensor operations (possible if $p_j$ is a trusted service), \system\ can block the handoff from $p_k$.  If so, the method to delay input events described above prevents such attacks.  However, \system\ may not know this mapping for third-party apps.

Thus, we extend the defense for input events to prevent ambiguity. When a target program is sent a handoff event associated with a user input event, then \system\ delays delivery of that event until the target program has completed processing any other handoff events.  Once the target program has begun processing a handoff associated with a user input event, \system\ prevents the delivery of subsequent handoff events until this processing completes.

Conceptually, this approach is analogous to placing a readers-writers lock over programs that may receive handoffs that result from user input events.  To avoid starving user input events (e.g., delaying them until the time limit), we prioritize delivery of handoffs that derive from user input events ahead of other handoffs using a simple, two-level scheduling approach.  We assess the impact of the proposed ambiguity prevention mechanisms on existing programs' functionality and performance in Section \ref{sec:performance}.

\subsection{Authorizing Delegation Paths}
\label{sec:authorization}

For controlling when a sensor operation may be performed as the result of a user input event, users are the only parties that know the intent of their actions. Therefore, users must be the party to make the final authorization decisions. To achieve this objective, \system\ elicits an explicit user authorization every time that a new delegation path is constructed (step~\circled{\scriptsize{\textbf{5}}} in
Figure~\ref{fig:strawman}).  Hence, to express a delegation path comprehensively, \system\ builds an authorization request that specifies: the user input event, all the programs that performed handoff events emanating from the user input event in the delegation path, the program performing the operation request, the operation to be performed, and the target sensor of the operation. As a result, all the programs that receive sensor data are clearly identified and reported in the authorization message, along with the input event, handoff events, and the resulting  sensor operation.

To reduce the user authorization effort, \system\ caches authorized delegation paths for reuse. After storing an authorized delegation path, \system\ proceeds in allowing the authorized sensor operation. For subsequent instances of the same user input event that results in exactly the same delegation path, which enables \system\ to omit step~\circled{\scriptsize{\textbf{5}}} and automatically authorize the sensor operation by leveraging the cached authorization.  Hence,  \system\ requires an explicit user's authorization \textit{only} the first time a delegation path is constructed for a specific user input event, similarly to the first-use permission approach.  As long as the program receiving a user input event does not change the way it processes that event (i.e., same handoffs), no further user authorization will be necessary.  In Section \ref{sec:user_study}, we show that such an approach does not prohibitively increase the number of access control decisions that users have to make thus avoiding decision fatigue \cite{Felt:2012}. 

Lastly, should a new delegation path be identified for a user input event that has a cached delegation path, then \system\ invalidates the cached delegation path and requires user authorization for the newly constructed delegation path before associating it to such an input event and caching it.

\section{Implementation}
\label{sec:implementation}

We implemented a prototype of the \system\ authorization system by modifying a recent release of the Android OS (Android-7.1.1\_r3) available via the Android Open Source Project (AOSP).\footnote{\scriptsize\texttt{https://source.android.com}} The choice of implementing a \system\ prototype for the Android OS was guided by its open-source nature and its wide adoption.  Its footprint is about 550 SLOC in C, 830 SLOC in C++, and 770 SLOC in Java. 

\begin{figure}[t]
\floatbox[{\capbeside\thisfloatsetup{capbesideposition={right,top},capbesidewidth=5cm, capbesidesep=quad}}]{figure}[\FBwidth]
{\caption{Authorization message prompted by \system\ to users the first time \textit{delegation paths} are created. Users are made aware of all the programs cooperating in serving their requests, as well as, of the entire delegation paths. Also, users are prompted with programs' names and identity marks to ease their identification. \system\ crosschecks developers' signatures and apps' identity (i.e., names and logos) by pulling information from the official Google Play store.}\label{fig:user-approval}}
{\includegraphics[width=3cm]{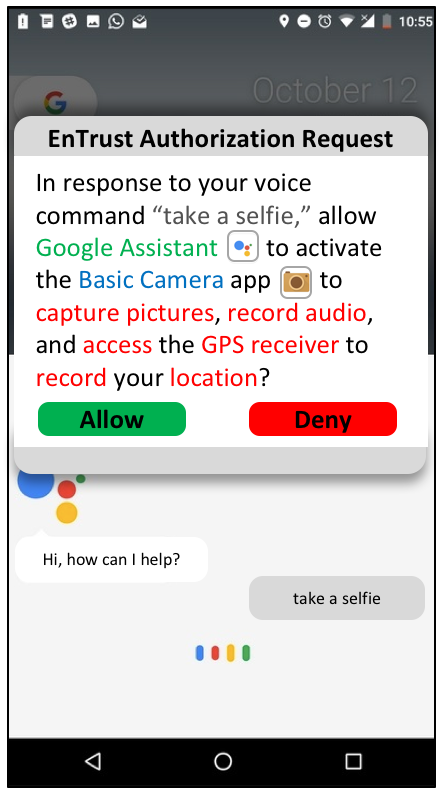}}
\end{figure}

\textbf{Program Identification} -- \label{sec:appid} To prove the programs' identity to users, \system\ specifies both the programs' name and visual identity mark (e.g., icon) in every {\em delegation request} as shown in Figure~\ref{fig:user-approval}. \system\ retrieves programs' identity by accessing the \texttt{AndroidManifest.xml}, which must contain a unique name and a unique identity mark (e.g., icon) for the program package. \system\ verifies programs' identity via the crypto-checksum\footnote{\scriptsize Android requires all apps and services to be signed by their developers.} of the program's binary signed with the
developer's private key and verifiable with the developer's public key \cite{signature}, similarly to proposed in prior work \cite{bianchi, petracca2017aware}. \system\ crosschecks developers' signatures and apps' identity (i.e., names and logos) by pulling information from the official Google Play store.\footnote{\scriptsize \texttt{https://play.google.com}} 

\textbf{User Input Event Authentication} -- \label{sec:user_input}
\system\ leverages SEAndroid \cite{smalley2013security}
to ensure that programs cannot inject input events by 
directly writing into input device files (i.e.,
\texttt{/dev/input/*}) corresponding to hardware and software input interfaces attached
to the mobile platform. Hence, only device drivers can write into input device files and only the Android Input Manager, a trusted system service, can read such device files and dispatch input events to programs.  Also, \system\ leverages the Android screen overlay mechanism to block overlay of graphical user interface components and prevent hijacking of user input events. Lastly, \system\ accepts only voice commands that are processed by the Android System Voice Actions module.\footnote{\scriptsize \texttt{https://developers.google.com/voice-actions/}}
 \system\ authenticates user input events by leveraging sixteen mediation hooks placed inside the stock Android Input Manager and six mediation hooks placed inside the System Voice Actions module.

\textbf{Handoff Event Mediation} -- Programs communicate with each other via Inter-Component Communication (ICC) that, in Android, is implemented as part of the Binder IPC mechanisms. The ICC includes both \textit{intent} and \textit{broadcast} messages that can be exchanged among programs. The Binder and the Activity Manager regulate messages exchanged among programs via the intent API.\footnote{\scriptsize \texttt{https://developer.android.com}} Programs can also send intents to other programs or services by using the broadcast mechanism that allows sending intents as arguments in broadcast messages. The Activity Manager routes intents to broadcast receivers based on the information contained in the intents and the broadcast receivers that have registered their interest in the first place. To mediate intents and broadcast messages exchanged between programs completely, \system\ leverages eight mediation hooks placed inside the Activity Manager and the Binder.

\textbf{Sensor Operation Mediation} -- \label{sec:mediation} Android uses the Hardware Abstraction Layer (HAL) interface to allow only
system services and privileged processes to access system
sensors indirectly via a well-defined API exposed by the
kernel. Moreover, SEAndroid~\cite{smalley2013security} is used to ensure that
only system services can communicate with the HAL at runtime. Any other program (e.g., apps) must interact with such system services to request execution of  operations targeting system sensors.
\system\ leverages such a mediation layer to identify operation requests generated by programs, by placing twelve hooks inside the stock
Android Audio System, Media Server, Location Services, and Media Projection.

\textbf{Event Scheduling} -- \label{sec:scheduling}
In Android, the Event Hub (part of the Input Manager server) reads raw input events from the input device driver files (\texttt{/dev/input/*}) and delivers them to the Input Reader. The Input Reader then formats the raw data and creates input event data that is delivered to the Input Dispatcher. The Input Dispatcher then consults the Window Manager to identify the target program based on the activity window currently displayed on the screen. Hence, we enhanced the Input Dispatcher to hold - for the duration of a time window - incoming input events for a target program should there be already a delivered input event for such a program that has not been processed yet. For handoff events, instead, the Binder is the single point of communication between two isolated programs trying to communicate view Inter-Process Communication (IPC). Therefore, the Binder has knowledge of all the pending messages exchanged between programs in the system and of the identity of the two communicating parties. Hence, we enhanced the Binder to hold - for the duration of a time window - incoming handoff events for a target program should the program be already involved in another communication with a third program.  Lastly, \system\ adopt a \textit{per-program} two-level queue scheduling, where events that can be tracked back to an input event have a higher priority. All other events are held in the lower priority queue. 

\textbf{Delegation Graph Management} -- \label{sec:request} The
\system\ prototype implements the \system\ {\sc Monitor} to handle
callbacks from the \system\ hooks inside the Input Manager, Activity Manager, Binder, and other
system services. The \system\ {\sc Monitor} is notified of mediated
events (e.g., input events, handoff events, and operation requests) via a
callback mechanism. Also, the \system\ {\sc Monitor} implements the logic for the \textit{delegation graph} construction, caching, and enforcement. 

\textbf{Authorization Message Management} --
The \system\ {\sc Monitor} displays \textit{authorization messages} to users for explicit authorizations of delegation paths, as shown in Figure \ref{fig:user-approval}. 
Also, \system\ prevents apps from creating windows that overlap the \system\ {\sc Monitor} messages by leveraging the Android screen overlay protection mechanism.
\system\ implements the \textit{Compartmented Mode Workstation} model \cite{compartmented}, by using
isolated per-window processes forked from the Window Manager, to prevent unauthorized modification of messages from the \system\ {\sc Monitor} by other programs.

\section{\system\ Evaluation} \label{sec:eval}

We investigated the following research questions:

$\blacktriangleright$ \textit{To what degree is the \system\ authorization assisting users in avoiding confused deputy, Trojan horse, and man-in-the-middle attacks?} We performed a \textit{laboratory-based} user study and found that  \system\ significantly increased  (from 54-64\% improvement) the ability of participants in avoiding attacks.

$\blacktriangleright$ \textit{What is the decision overhead imposed by \system\ on users due to explicit authorization of constructed delegation graphs?} We performed a \textit{field-based} user study and found that the number of decisions imposed on users by \system\ remained confined to no more than 4 explicit authorizations per program. 

$\blacktriangleright$ \textit{Is \system\ backward compatible with existing programs? How many operations from legitimate programs are incorrectly blocked by \system?} 
We used a well-known compatibility test suite to evaluate the compatibility of \system\ with 1,000 apps (selected among the most popular apps on Google Play) and found that \system\ does not cause the failure of any application.

$\blacktriangleright$ \textit{What is the performance overhead imposed by \system\ for the delegation graph construction and enforcement?} We used a well-known software exerciser to measure the performance overhead imposed by \system. We found that \system\ introduced a negligible overhead (order of milliseconds) unlikely noticeable to users.

\begin{table*}[]
\centering
\setlength{\tabcolsep}{.05em}
\scriptsize
\caption{
Experimental tasks for the laboratory-based user study, derived from the attack vectors described in Section~\ref{sec:problem}. We report the delegation graphs constructed by \system\ and the authorization messages presented to subjects in the four groups. Authorizations request prompted by \system\ include programs' identity marks, as shown in Figure~\ref{fig:user-approval}, omitted in this Table.} 
\label{tab:tasks}

\begin{tabular}{l|p{1.45cm}|p{2.8cm}|p{3.95cm}|p{4.8cm}|p{4.8cm}|p{4.8cm}|p{4.8cm}|}
\cline{2-6}
 & \multicolumn{1}{c|}{\cellcolor[HTML]{C0C0C0}\textbf{Task Directive}} & \multicolumn{1}{c|}{\cellcolor[HTML]{C0C0C0}\textbf{Attack Scenario}} & \multicolumn{1}{c|}{\cellcolor[HTML]{C0C0C0}\textbf{Delegation Graph}} & \multicolumn{2}{c|}{\cellcolor[HTML]{C0C0C0}\textbf{First-Use}} & \multicolumn{2}{c|}{\cellcolor[HTML]{C0C0C0}\textbf{EnTrust}} \\ \hline

\multicolumn{1}{|c|}{\cellcolor[HTML]{EFEFEF}\begin{tabular}[t]{@{}c@{}}\textbf{T}\\\textbf{A}\\\textbf{S}\\\textbf{K}\\\squared{\scriptsize{\textbf{A}}} \end{tabular}} 

& \raggedright Ask \textbf{Smart Assistant} to ``create a note.'' Dictate a voice note to \textbf{Notes}. 

& \raggedright \underline{Confused Deputy}: \textbf{Smart Assistant} opens the \texttt{Notes} app and adds the specified note, however, it also sends a request to the \textbf{Screen Capture} service to capture the content on the screen. A note containing the credit card information of a purchasing card is captured from the \textbf{Notes} app and sent to a remote server.  
&  \raisebox{-\totalheight}{\includegraphics[width=40mm]{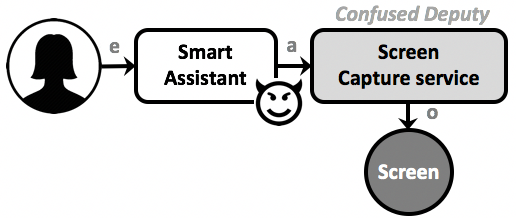}} 

& \multicolumn{2}{p{4.8cm}|}{\raggedright \textit{\textbf{Smart Assistant} will start \textbf{capturing} everything that's displayed on your \textbf{screen}.} \hfill \newline \hfill \newline \hfill \newline  \hfill \newline \hfill  \newline \hfill \newline \hfill \newline \hfill
\begin{tabular}[c]{p{2.5cm} |p{2.3cm}}
\hline
\cellcolor[HTML]{EFEFEF} \textbf{{\bluemarkA{\texttt{Group-FR-U}}}(nprimed)} & \cellcolor[HTML]{EFEFEF} \textbf{{\bluemarkB{\texttt{Group-FR-P}}}(rimed)} \\
\textbf{\color{red} 81\% Attack Success} & \textbf{\color{red} 55\% Attack Success}  \\
\textbf{\color{black!30!brown} 55\% Prompted} & \textbf{\color{black!30!brown} 45\% Prompted}\\
\textbf{\color{black!30!orange}36\% Explicit Allows} & \textbf{\color{black!30!orange}0\% Explicit Allows}\\
\end{tabular}}

& \multicolumn{2}{p{4.8cm}|}{\raggedright In response to your voice command ``create a note'', allow  \textbf{Smart Assistant} to activate the \textbf{Screen Capture} service to \textbf{capture} the content on the \textbf{screen}? \newline \hfill \newline \hfill \newline \hfill \newline \hfill \newline \hfill
\begin{tabular}[c]{p{2.5cm} |p{2.2cm}}
\hline
\cellcolor[HTML]{EFEFEF} \textbf{{\yellowmarkA{\texttt{Group-EN-U}}}(nprimed)} & \cellcolor[HTML]{EFEFEF} \textbf{{\yellowmarkB{\texttt{Group-EN-P}}}(rimed)} \\
\textbf{\color{red} 27\% Attack Success} & \textbf{\color{black!30!green} 0\% Attack Success}\\ 
\textbf{\color{black!30!brown} 100\% Prompted} & \textbf{\color{black!30!brown} 100\% Prompted}\\
\textbf{\color{black!30!orange}27\% Explicit Allows} & \textbf{\color{black!30!orange}0\% Explicit Allows}\\
\end{tabular}} \\ \hline

\multicolumn{1}{|c|}{\cellcolor[HTML]{EFEFEF}\begin{tabular}[t]{@{}c@{}}\textbf{T}\\\textbf{A}\\\textbf{S}\\\textbf{K}\\\squared{\scriptsize{\textbf{B}}} \end{tabular}} 

& \raggedright Ask \textbf{Google Assistant} to ``take a selfie.'' 

& \raggedright \underline{Trojan Horse}: \textbf{Google Assistant} activates the \textbf{Basic Camera} app, which is a Trojan app that takes a selfie but also records a short audio and the user's location. The collected data is then sent to a remote server accessible by the adversary. 
&  \raisebox{-\totalheight}{\includegraphics[width=40mm]{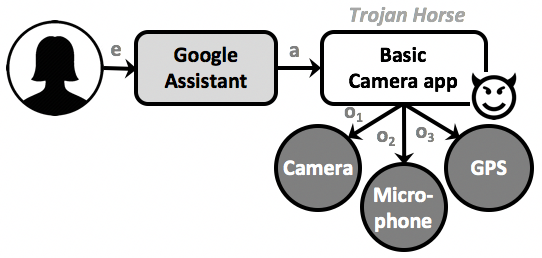}} 

& \multicolumn{2}{p{4.8cm}|}{\raggedright \textit{Allow \textbf{Basic Camera} to \textbf{capture pictures}?} \\ \textit{Allow \textbf{Basic Camera} to \textbf{record audio?}} \\ \textit{Allow \textbf{Basic Camera} to \textbf{access} this device's \textbf{location?} \newline \hfill \newline \hfill \newline \hfill }
\begin{tabular}[c]{p{2.5cm} |p{2.3cm}}
\hline
\cellcolor[HTML]{EFEFEF} \textbf{{\bluemarkA{\texttt{Group-FR-U}}}(nprimed)} & \cellcolor[HTML]{EFEFEF} \textbf{{\bluemarkB{\texttt{Group-FR-P}}}(rimed))} \\
\textbf{\color{red} 81\% Attack Success} & \textbf{\color{red} 55\% Attack Success}\\
\textbf{\color{black!30!brown} 36\% Prompted} & \textbf{\color{black!30!brown} 45\% Prompted}\\
\textbf{\color{black!30!orange}18\% Explicit Allows} & \textbf{\color{black!30!orange}0\% Explicit Allows}\\
\end{tabular}}

& \multicolumn{2}{p{4.8cm}|}{\raggedright In response to your voice command ``take a selfie'', allow \textbf{Google Assistant} to activate the \textbf{Basic Camera} app to \textbf{capture pictures}, \textbf{record audio}, and access the \textbf{GPS receiver} to \textbf{record} your \textbf{location}? \newline \hfill  \newline \hfill
\begin{tabular}[c]{p{2.5cm} |p{2.2cm}}
\hline
\cellcolor[HTML]{EFEFEF} \textbf{{\yellowmarkA{\texttt{Group-EN-U}}}(nprimed)} & \cellcolor[HTML]{EFEFEF} \textbf{{\yellowmarkB{\texttt{Group-EN-P}}}(rimed)} \\
\textbf{\color{red} 18\% Attack Success} & \textbf{\color{black!30!green} 0\% Attack Success}\\
\textbf{\color{black!30!brown} 100\% Prompted} & \textbf{\color{black!30!brown} 100\% Prompted}\\
\textbf{\color{black!30!orange}18\% Explicit Allows} & \textbf{\color{black!30!orange}0\% Explicit Allows}\\
\end{tabular}} \\ \hline

\multicolumn{1}{|c|}{\cellcolor[HTML]{EFEFEF}\begin{tabular}[t]{@{}c@{}}\textbf{T}\\\textbf{A}\\\textbf{S}\\\textbf{K}\\\squared{\scriptsize{\textbf{C}}} \end{tabular}} 

& \raggedright Ask \textbf{Google Assistant} to ``deposit \textit{bank} check.'' After logging into \textbf{Mobile Banking} with the provided credentials, deposit the provided check.
& \raggedright \underline{Man-In-The-Middle}: \textbf{Google Assistant} launches \textbf{Basic Camera} registered for the voice intent ``deposit \textit{bank} check''. The \textbf{Basic Camera} runs in the background, captures a picture of the check and - via a spoofed intent - launches the \textbf{Mobile Banking} app registered for the voice intent ``deposit check''. 
&  \raisebox{-\totalheight}{\includegraphics[width=40mm]{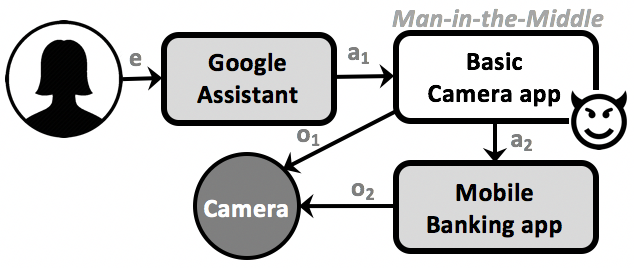}} 

& \multicolumn{2}{p{4.8cm}|}{\raggedright \textit{Allow \textbf{Basic Camera} to \textbf{capture pictures}?} \\\textit{Allow \textbf{Mobile Banking} to \textbf{capture pictures}?}
\newline \hfill \newline \hfill \newline \hfill \newline \hfill \newline \hfill \newline \hfill \newline \hfill \newline \hfill
\begin{tabular}[c]{p{2.5cm} |p{2.3cm}}
\hline
\cellcolor[HTML]{EFEFEF} \textbf{{\bluemarkA{\texttt{Group-FR-U}}}(nprimed)}  & \cellcolor[HTML]{EFEFEF} \textbf{{\bluemarkB{\texttt{Group-FR-P}}}(rimed)} \\
\textbf{\color{red} 73\% Attack Success} & \textbf{\color{red} 55\% Attack Success}\\
\textbf{\color{black!30!brown} 45\% Prompted} & \textbf{\color{black!30!brown} 45\% Prompted}\\
\textbf{\color{black!30!orange}18\% Explicit Allows} & \textbf{\color{black!30!orange}0\% Explicit Allows}\\
\end{tabular}}

& \multicolumn{2}{p{4.8cm}|}{\raggedright In response to your voice command ``deposit a check,'' allow \textbf{Google Assistant} to activate the \textbf{Basic Camera} app to \textbf{capture pictures}.  Also, allow  the \textbf{Basic Camera} app to activate the \textbf{Mobile Banking} app to \textbf{capture pictures}? \newline \hfill \newline \hfill \newline \hfill \newline \hfill \newline \hfill
\begin{tabular}[c]{p{2.5cm} |p{2.2cm}}
\hline
\cellcolor[HTML]{EFEFEF} \textbf{{\yellowmarkA{\texttt{Group-EN-U}}}(nprimed)}  & \cellcolor[HTML]{EFEFEF} \textbf{{\yellowmarkB{\texttt{Group-EN-P}}}(rimed)} \\
\textbf{\color{red} 9\% Attack Success} & \textbf{\color{black!30!green} 0\% Attack Success}\\
\textbf{\color{black!30!brown} 100\% Prompted} & \textbf{\color{black!30!brown} 100\% Prompted}\\
\textbf{\color{black!30!orange} 9\% Explicit Allows} & \textbf{\color{black!30!orange}0\% Explicit Allows}\\
\end{tabular}} \\ \hline
\end{tabular}
\end{table*}

\subsection{User Study Preliminaries}

We designed our user studies following suggested practices for human subject studies in security to avoid common pitfalls in conducting and writing about security and privacy human subject research \cite{common-pitfalls}. Our study is comprehensive, however, does not focus explicitly on long-term habituation. In particular, researchers have studied user's habituation for first-use authorization systems extensively \cite{Felt:2012, primal} and our field-based study (Section \ref{subsec:field-based-study}) shows that our approach is comparable to first-use in terms of the number of times users are prompted. Therefore, we expect that results reported in prior work should apply to our approach as well \cite{Felt:2012, primal}. Further, researchers have extensively studied peoples' attitudes toward privacy \cite{Petracca2016AwareCA, sheehan2002toward, debatin2009facebook}, therefore, we did not find it interesting to measure users' attitude toward privacy once again during our studies. An Institutional Review Board (IRB) approval was obtained from our institution. We recruited user study participants via local mailing lists, Craigslist, Twitter, and local groups on Facebook. We compensated them with a \$5 gift card.  We excluded acquaintances from participating in the studies to avoid acquiescence bias. We made sure to get a wide diversity of subjects, both in terms of age and experience with technology. Participants were given the option to withdraw their consent to participate at any time after the purpose of the study was revealed. For all the experiments, we configured the test environment on LG Google Nexus 5X phones running the Android 7.1 Nougat OS. We used a background service, automatically relaunched at boot time, to log participants' responses to system messages and alerts, as well as all user input events generated by participants while interacting with the testing programs.  During the experiments, the researchers took note of comments made by participants.

\subsubsection{\textbf{Laboratory-Based User Study}} \label{sec:user_study}
\hfill

We performed a \textit{laboratory-based} user study to evaluate the effectiveness of \system\ in supporting users in avoiding all the three attack vectors previously identified in Section~\ref{sec:problem}. We compared \system\ with the \textit{first-use} authorization used in commercial systems. We could not compare mechanisms proposed in related work \cite{udac, udac2, petracca2017aware}, because they are unable to handle handoff events. We divided participants into four groups, participants in {\bluemarkA{{\bluemarkA{\texttt{Group-FR-U}}}}} and {\bluemarkB{{\bluemarkB{\texttt{Group-FR-P}}}}} interacted with a \textit{stock} Android OS implementing the \textit{first-use} authorization mechanism. Participants in {\yellowmarkA{{\yellowmarkA{\texttt{Group-EN-U}}}}} and {\yellowmarkB{{\yellowmarkB{\texttt{Group-EN-P}}}}} interacted with a \textit{modified} version of the Android OS integrating the \system\ authorization system. To account for the priming effect, we avoided influencing subjects in  {\texttt{Group-FR-U}}  and  {\texttt{Group-EN-U}}  and advertised the test as a generic ``voice assistants testing'' study without mentioning security implications. On the other hand, to assess the impact of priming, subjects in  {\texttt{Group-FR-P}} and  {\texttt{Group-EN-P}} were informed that attacks targeting system sensors (e.g., camera, microphone, and GPS receiver) were possible during the interaction with programs involved in the experimental tasks, but without specifying what program performed the attacks or what attacks were performed. 

\underline{\textit{Experimental Procedures}}:  For our experiments, we used two voice assistant programs. The Smart Assistant is a test assistant developed in our research lab that provides basic virtual assistant functionality, such as voice search, message composition, and note keeping. However, Smart Assistant is also designed to perform confused deputy attacks on system services, such as the Screen Capture service. The Google Assistant, instead, is the benign assistant shipped with the stock Android OS. We also used a test app, Basic Camera, developed in our research lab. It provides basic camera functionality, such as capturing pictures or videos and applying photographic filters. However, Basic Camera is also designed to perform man-in-the-middle and Trojan horse attacks for requests to capture photographic frames. Lastly, we used a legitimate Mobile Banking app (from a major international bank) available on Google Play. 

All the instructions to perform the experimental tasks were provided to participants via a handout at the beginning of the user study. Before starting the experimental tasks, we asked the participants to familiarize themselves with the provided smartphone, by interacting with the voice assistants and other available programs. We will refer to this phase as the \textit{preliminary phase}. This phase is meant to avoid a ``cold start'' and approximate a scenario in which the users have been using their device. \textit{During this phase, all the participants were prompted with authorization requests at first-use of any of the system sensors.} Afterwards, all participants were asked to perform, in a randomized order, the three experimental tasks reported in Table \ref{tab:tasks}. Each participant was presented with the corresponding authorization messages reported in Table \ref{tab:tasks} depending on the group to which they were assigned.\footnote{\scriptsize The runtime permission mechanism enabled subjects to revoke granted permissions at any time.} 

\underline{\textit{Experimental Results}}: In total, forty-four subjects participated in
and completed our laboratory-based user study. We \textit{randomly} assigned 11
participants to each group. Table~\ref{tab:tasks}
summarizes the results of the three experimental tasks. Overall, we found that the delegation graphs constructed by \system\ aided the participants in avoiding attacks. \system\ significantly outperformed  the \textit{first-use} approach currently used in commercial systems.  In fact, with the first-use approach  participants may not have been prompted with authorization messages once again during the experimental tasks, should they have authorized the program in a previous task or during the preliminary phase as per first-use default. Instead, with \system\ a new authorization message was presented to the participants whenever a new delegation path was identified, thus, making the delegation explicit to users. This explains the lower attack success rates relative to the first-use groups. 

{{\scriptsize\bf{\marked{TASK}\squared{A}}}}: The analysis of subjects' responses revealed that 5 subjects from {\texttt{Group-FR-U}} and 6 subjects from {\texttt{Group-FR-P}} had interacted with Smart Assistant during the preliminary phase to ``take a screenshot'' and had granted the app permission to capture the their screen. Thus, they were not prompted once again with an authorization message during Task A, as per default in first-use permissions. In addition, 4 subjects from {\texttt{Group-FR-U}}  explicitly allowed Smart Assistant to capture their screen, therefore, resulting in a 81\% and 55\% attack success, respectively, as reported in Table~\ref{tab:tasks}. On the contrary, only 3 subjects from {\texttt{Group-EN-U}} and no subjects from {\texttt{Group-EN-P}} allowed the attack (27\% and 0\% attack success, respectively). Also, similarly to what happened to participants in {\texttt{Group-FR-U}} and {\texttt{Group-FR-P}}, 6 subjects from {\texttt{Group-EN-U}} and 6 subjects from {\texttt{Group-EN-P}} had interacted with Smart Assistant during the preliminary phase and asked to ``take a screenshot.'' However, since the voice command ``create a note'' was a different command, \system\ prompted all the participants with a new authorization message, as shown in Table~\ref{tab:tasks}. 

{{\scriptsize\bf{\marked{TASK}\squared{B}}}}: The analysis of subjects' responses revealed that  7 subjects from {\texttt{Group-FR-U}} and 6 subjects from {\texttt{Group-FR-P}} had interacted with Basic Camera to take a picture or record a video, either during the preliminary phase or during another task, and authorized it to capture pictures, audio, and access the device's location. Thus, they were not prompted once again during this task as per default in first-use permissions. Also, we found that 2 subjects from {\texttt{Group-FR-U}} explicitly authorized Basic Camera to access the camera, as well as the microphone, and the GPS receiver; therefore, resulting in a 81\% and 55\% attack success, respectively.  In contrast, 2 subjects from {\texttt{Group-EN-U}} and no subject from  {\texttt{Group-EN-P}} authorized access to the camera, microphone, and GPS receiver (18\% and 0\% attack success, respectively). Also, we found that 8 subjects from {\texttt{Group-EN-U}} and 6 subjects from {\texttt{Group-EN-P}} had interacted with Basic Camera during the preliminary phase or during another task. However, none of them asked to ``take a selfie'' before, so all participants were prompted by \system\ with a new authorization message. At the end of the experiment, among all the participants, when asked why they had authorized access to the GPS receiver the majority said that they expected a camera app to access location to create geo-tag metadata while taking a picture. In contrast, for those who had denied the permission, their reasoning was that they did not feel comfortable sharing their location \textit{when taking a selfie}.

{{\scriptsize\bf{\marked{TASK}\squared{C}}}}: The analysis of subjects' responses revealed that 6 subjects form {\texttt{Group-FR-U}} and 6 subjects from {\texttt{Group-FR-P}} had interacted with Basic Camera either during the preliminary phase or during another task and authorized the app to capture pictures. Thus, during this task, they were not prompted with an authorization message once again as per default in first-use permissions. They were only prompted to grant permission to Mobile Banking, explaining why even the primed subjects were not able to detect the attack. 
In addition, 2 subjects from {\texttt{Group-FR-U}} explicitly authorized Basic Camera to capture a frame with the bank check; therefore, resulting in a 73\% and 55\% attack success, respectively. On the other hand, only 1 subject from {\texttt{Group-EN-U}} and no subjects from {\texttt{Group-EN-P}} had authorized Basic Camera to capture a frame with the bank check, resulting in a 9\% and 0\% attack success, respectively. Notice that all the participants from {\texttt{Group-EN-U}} and {\texttt{Group-EN-P}} were prompted with a new authorization message by \system\ for the new command ``deposit \textit{bank} check.'' Interestingly, the one subject from {\texttt{Group-EN-U}}, who had allowed Basic Camera to capture a frame with the bank check, verbally expressed his concern about the permission notification presented on the screen. The subject stated observing that two apps asked permission to access the camera to take pictures. This is not unreasonable for an unprimed participant, who does not expect a malicious behavior.

\underline{\textit{Discussion}}: By comparing the results from {\texttt{Group-FR-U}} versus those from {\texttt{Group-FR-P}}, and those from {\texttt{Group-EN-U}} versus those from {\texttt{Group-EN-P}}, we observed that primed subjects allowed fewer attacks. However, \system\ was significantly more effective than first-use in keeping users ``on guard'' independently of whether participants were primed (54-64\% lower attack success with \system). Indeed, differently from first-use or prior defense mechanisms \cite{petracca2017aware, udac, udac2}, \system\ was able to identify whether pre-authorized programs attempted accessing system sensors via unauthorized delegation paths, which could potentially had affected users' privacy and security. If so, \system\ prompted users for an explicit authorization of newly identified delegation paths.  Also, \system\ performed slightly better than first-use authorization for explicit user authorizations (Explicit Allows in Table~\ref{tab:tasks}).  Thus, additional information provided by \system\ in authorization messages (i.e., programs' name and identity mark, as well as, delegation information, as shown in Figure~\ref{fig:user-approval}), appears to be helpful to users in avoiding unexpected behaviors from programs.

\subsubsection{\textbf{Field-Based User Study}}
\label{subsec:field-based-study}
\hfill

We performed a \textit{field-based} user study to evaluate whether \system\ increases the decision-overhead imposed on users. We measured the number of explicit authorizations users had to make when interacting with \system\ under realistic and practical conditions, and compared it with the first-use approach adopted in commercial systems. We also measured the number of authorizations handled by \system\ via the cache mechanism that, transparently to users, granted authorized operations.

\underline{\textit{Experimental Procedures}}: Participants were provided with an LG Nexus 5X smartphone running a \textit{modified} version of the Android OS integrating the \system\ authorization framework, and asked to use it for 7 days. During this period, participants interacted with 5 voice assistants and 10 apps selected among the most popular\footnote{\scriptsize Source: \scriptsize\texttt{https://fortune.com}} with up to millions of downloads from the official Google Play store.  Participants were asked to explore each voice assistant and app at least once by interacting as they would normally do. Particularly, we asked the participants to interact with each voice assistant by asking the following three questions: (1) ``capture a screenshot,'' (2) ``record a voice note,'' (3) ``how long does it take to drive back home.'' Additionally, we asked participants to be creative and ask three additional questions of their choice. Mock accounts were provided to participants for apps requiring a log-in. Table \ref{tab:tested-apps} summarizes all the assistants and apps pre-installed on the smartphones for the field-based user study. The smartphones provided to participants were running a background service with runtime logging enabled, automatically restarted at boot time, to monitor the number of times each program was launched, the users' input events, the constructed delegation graphs, the authorization decisions made by the participants, and the number of authorizations automatically granted by \system. The background service also measured the gaps between consecutive input events and handoff events, as well as the time required by each program to service each event.

\begin{table}[]
\centering
\scriptsize
\begin{tabular}{l|c|c|
>{\columncolor[HTML]{9B9B9B}}c |}
\cline{2-4}
 & \multicolumn{2}{c|}{\cellcolor[HTML]{C0C0C0}Explicit Authorizations} & \multicolumn{1}{c|}{\cellcolor[HTML]{9B9B9B}} \\ \cline{2-3}
 & \multicolumn{1}{c|}{\cellcolor[HTML]{C0C0C0}\texttt{First-Use}} & \multicolumn{1}{c|}{\cellcolor[HTML]{C0C0C0}\system} & \multicolumn{1}{c|}{\multirow{-2}{*}{\cellcolor[HTML]{9B9B9B}\begin{tabular}[c]{@{}c@{}}Authorized\\ Operations (7 days)\end{tabular}}} \\ \hline
\multicolumn{1}{|l|}{\cellcolor[HTML]{EFEFEF}\begin{tabular}[c]{@{}l@{}}Snapchat\\YouTube\\Facebook Messenger\\ Instagram\\Facebook\\Whatsapp\\Skype\\WeChat\\Reddit\\Bitmoji\end{tabular}} & \begin{tabular}[c]{@{}c@{}}3\\3\\2\\3\\3\\2\\3\\2\\1\\3\end{tabular} & \begin{tabular}[c]{@{}c@{}}3\\3\\2\\3\\3\\2\\3\\2\\1\\3\end{tabular} &\begin{tabular}[c]{@{}c@{}}276\\84\\93\\393\\117\\76\\100\\101\\18\\127\end{tabular}\cellcolor[HTML]{EFEFEF} \\ \hline
\multicolumn{1}{|l|}{\cellcolor[HTML]{EFEFEF}\begin{tabular}[c]{@{}l@{}}Google Assistant\\Microsoft Cortana\\ Amazon Alexa\\Samsung Bixby\\Lyra Virtual Assistant\end{tabular}} & \begin{tabular}[c]{@{}c@{}}1\\1\\1\\1\\1\end{tabular} & \begin{tabular}[c]{@{}c@{}}\textbf{4}\\\textbf{3}\\\textbf{4}\\\textbf{4}\\\textbf{3}\end{tabular}  & \cellcolor[HTML]{EFEFEF}\begin{tabular}[c]{@{}c@{}}72\\49\\84\\63\\56\end{tabular}\\ \hline
\end{tabular}
\caption{Apps and voice assistants tested in the field study. The last column shows the number of operations automatically authorized by \system\ after user's authorization.}
\label{tab:tested-apps}
\end{table}

\underline{\textit{Experimental Results}}: Nine subjects participated and completed the field-based user study. The data collected during our experiment indicates that all user authorizations were obtained within the first 72 hours of interaction with the experimental device, after which we observed only operations automatically granted by \system.

The first participant allowed us to discover two implementation issues that affected the number of explicit authorizations required by \system. First, changing the orientation of the screen (portrait versus landscape) was causing \system\ to request a new explicit user authorization for an already authorized widget whenever the screen orientation changed. This inconvenience was due to the change in some of the features used to model the context within which the widget was presented. To address this shortcoming, we modified our original prototype to force the Window Manager to generate in memory two graphical user interfaces for both screen orientations to allow \system\ to bind them with a specific widget presented on the screen. Second, for the voice commands, we noticed that differently phrased voice commands with the same meaning would be identified as different user input events. For instance, ``take a selfie'' and ``take a picture of me''. This shortcoming was causing \system\ to generate a new delegation graph for each differently phrased voice command. To address this issue, we leveraged the \textit{Dialogflow} engine by Google, part of the AI.API.\footnote{\scriptsize\texttt{ https://dialogflow.com}} \textit{Dialogflow} is a development suite for building conversational interfaces and provides a database of synonyms to group together voice commands with the same meaning. We fixed the two issues and continued our experiment with the other participants.

Table~\ref{tab:tested-apps} reports the average number of explicit authorizations performed by the participants. 
We compared them with the number of explicit authorizations that would be necessary if the \textit{first-use} permission mechanism was used instead. The results show that \system\ required the same number of explicit authorizations by users for all the tested apps. For all voice assistants, instead, \system\ may required up to 3 additional explicit authorizations when compared with the \textit{first-use} approach. These additional authorizations are due to the fact that with the \textit{first-use} approach the programs activated by the voice assistant to serve the user request may have already received the required permission to access the sensitive sensors. \system\ instead captures the entire sequence of events, from the user input event to any subsequent action or operation request, and then ties them together. Therefore, \system\ constructs a new graph for each different interaction.  Nonetheless, the number of decisions imposed on the users remains very modest. Indeed, on average, three additional explicit user authorizations are required per each voice assistant. Also, the number of explicit authorizations made by the users remained a constant factor compared to the number of automatically authorized operations, which instead grew linearly over time. We measured an average of 16 operations automatically authorized by \system\ during a 24-hour period (last column of Table~\ref{tab:tested-apps}). Therefore, if we consider the daily average number of automatically authorized operations for a period of one year, we will have on the order of thousands of operations automatically authorized by \system, which would not require any additional explicit effort for the users.

\begin{figure}[t]
\centering
\includegraphics[width=90mm]{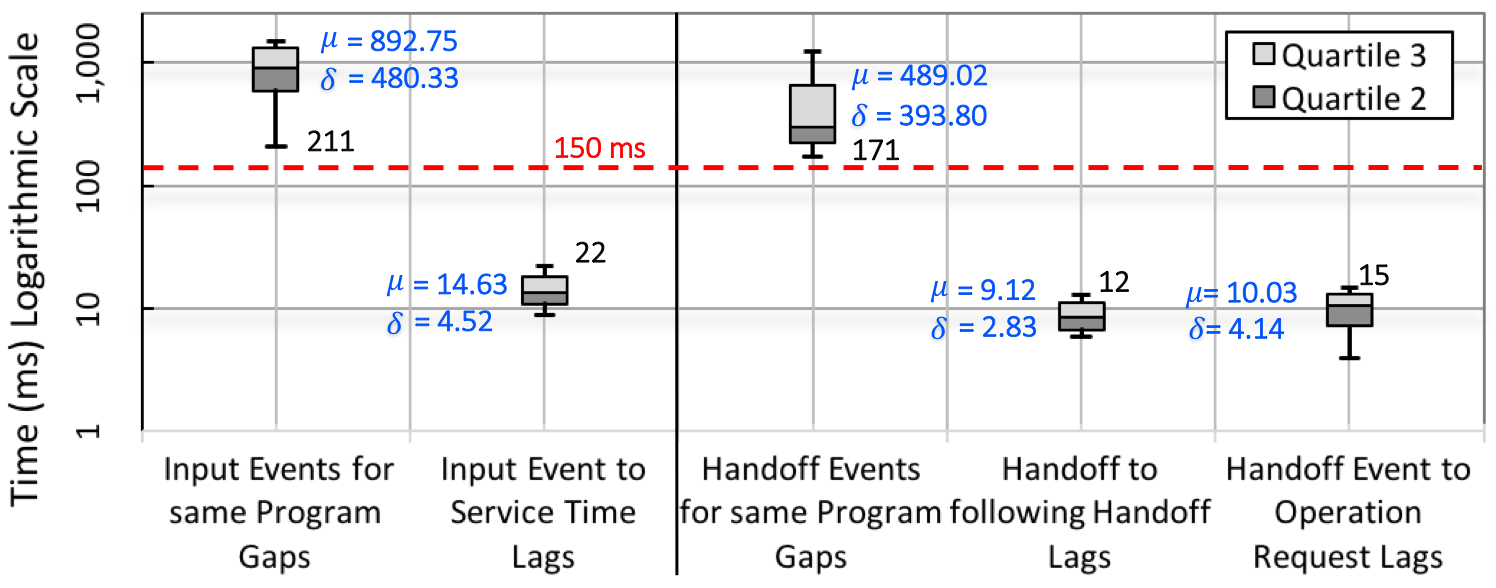}
\caption{Time analysis used to study the possibility of ambiguous events delegation paths, as discussed in Section~\ref{sec:no_ambiguous}.}
\label{fig:time-conditions}
\end{figure}

\begin{figure*}[t]
\setlength{\tabcolsep}{.3em}
\centering
\begin{tabular}{cccc}
  \includegraphics[width=44mm]{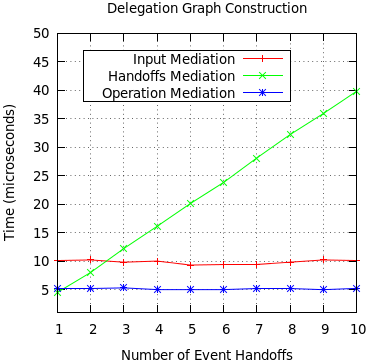} &  \includegraphics[width=45mm]{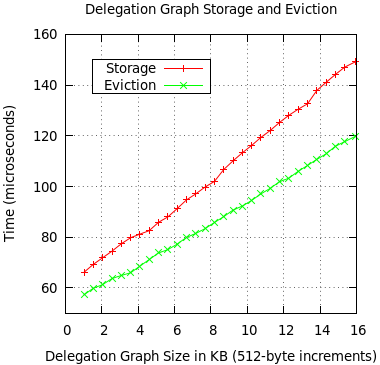} &  \includegraphics[width=44mm]{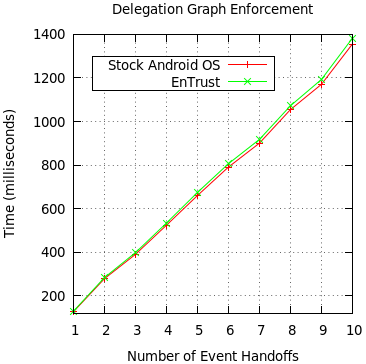} &  \includegraphics[width=42mm]{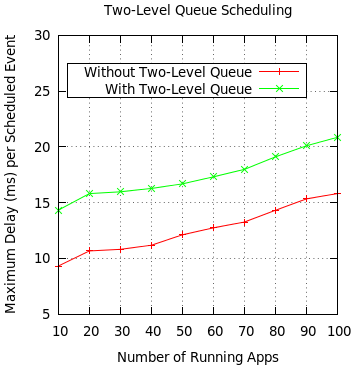}  
\\
\end{tabular}
\caption{Overheads for Delegation Graphs Construction, Storage, Eviction, Enforcement, and Two-Level Queue Scheduling.}\label{fig:benchmarks}
\end{figure*}

\underline{\textit{Time Constraints Analysis}}:
\label{sec:time-analysis} We leveraged data collected via the field-based user study to perform an analysis of time constraints for input events and action/operation requests to calibrate the time window for the event ambiguity prevention mechanism (Section~\ref{sec:no_ambiguous}). Figure~\ref{fig:time-conditions} reports the measurements of the gaps\footnote{\scriptsize Gaps higher than 1,500 ms were excluded because not relevant to the analysis.} between consecutive input events and consecutive handoff events, as well as the lags between each event and the corresponding response from the serving program. From the measurements, we observed: (1) the minimum gap between subsequent input events targeting the same program (211 ms) is an order of magnitude larger than the maximum lag required by the program to serve each incoming event (22 ms); and (2) the minimum gap (171 ms) between subsequent handoff events targeting the same program is an order of magnitude larger than the maximum lag required by the program to serve incoming requests (15 ms). Hence, to avoid ambiguity, we may set the time window to 150 ms to guarantee that the entire delegation path can be identified before the next event for the same program arrives. Lastly, we observed that 87\% of the delegation paths had a total length of three edges (one input event, one handoff event, and one sensor operation request). The remaining 13\% of the delegation paths had a maximum length of four edges (one additional handoff event), which further supports our claim that we can hold events without penalizing concurrency of such events. 

\subsection{Backward Compatibility Analysis} \label{sec:compatibility}

To verify that \system\ is backward compatible with existing programs, we used the Compatibility Test Suite (CTS),\footnote{\scriptsize\texttt{https://source.android.com/compatibility/cts/}} an automated testing tool released by Google via the AOSP.\footnote{\scriptsize\texttt{Android Open Source Project - https://source.android.com}} In particular, this analysis verified that possible delays in the delivery of events introduced by \system\ or the change in scheduling of events did not impact applications' functionality. We tested the compatibility of \system\ with 1,000 existing apps, among the top 2,000 most downloaded apps on Google Play, selected based on those declaring permissions to access sensitive sensors in their manifest. The experiment took 19 hours and 45 minutes to complete, and \system\ passed 132,681 tests without crashing the operating system and without incorrectly blocking any legitimate operation. Among the 1,000 tested apps, we also included 5  popular augmented reality multi-player gaming app (\texttt{InGress}, \texttt{Pok\'emon Go}, \texttt{Parallel Kingdom}, \texttt{Run An Empire}, and \texttt{Father.io}), which typically have a high rate of input events and are very sensitive to delays. The set of tests targeting these 5 gaming apps ran for 16 minutes, during which we continuously observed the device screen to identify possible issues in terms of responsiveness to input events or glitches in the rendering of virtual objects on the screen. However, we did not identify any discernible slowdown, glitch, or responsiveness issue.

\subsection{Performance Measurements}
\label{sec:performance}
We performed five micro-benchmarks on a standard Android developer smartphone, the LG Nexus 5X, powered by 1.8GHz hexa-core 64-bit Qualcomm Snapdragon 808 Processor and Adreno 418 GPU, 2GB of RAM, and 16GB of internal storage. 
All of our benchmarks are measured using Android 7.1 Nougat pulled from the Android Open Source Project (AOSP) repository on February 21, 2017.  

\textbf{Delegation Graph Construction} -- Our first micro-benchmark of \system\ measured the overhead incurred for constructing delegation graphs of varying sizes. To do this, we had several programs interacting and generating a handoff-events chain varying from 1 to 10 handoffs in length and measured the time to mediate the input event, the handoff event, and the operation request. We repeated the measurements 100 times. Each set of measurements was preceded by a priming run to remove any first-run effects. We then took an average of the middle 8 out of 10 such runs for each number of handoff events. 
The results in Figure~\ref{fig:benchmarks} show that the input mediation requires an overhead of 10 $\mu$s, the handoff event mediation requires an additional overhead of 4 $\mu$s per event handoff, whereas the operation mediation requires a fixed overhead of 5 $\mu$s. The overheads are within our expectations and do not cause noticeable performance degradation. 

\textbf{Delegation Graph Caching} -- Our second micro-benchmark of \system\ measures the overhead incurred for caching delegation graphs constructed at runtime. These results are intended to measure the overhead introduced by \system\ in the authorization process for both storing a new delegation graph, as well as evicting from cache a stale one. To do this, we simulated the creation and eviction of delegation graphs of different sizes varying from 1 to 16 Kilobytes in 512-byte increments.\footnote{\scriptsize This range was selected based on the size of the delegation graphs created during our experiments, which should be representative of real scenarios.} We repeated the measurement 5 times for each random size and took an average of the middle 3 out of 5 such runs. The results in Figure~\ref{fig:benchmarks} show that the storing of delegation graphs in the cache required a base overhead of 66 $\mu$s with an additional 3 $\mu$s per 512-byte increment. The eviction instead required a base overhead of 57 $\mu$s with an additional 2.5 $\mu$s for each 512-byte increment.

\textbf{Delegation Graph Enforcement} -- Our third micro-benchmark was designed to compare the unmodified version of the Android Nougat build for control measurement with a modified build integrating our \system\ features for the delegation graph enforcement during authorization.  To guarantee fairness in the comparison between the two systems, we used the Android UI/Application Exerciser Monkey\footnote{\scriptsize\texttt{https://developer.android.com/studio/test/monkey.html}} to generate the same sequence of events for the same set of programs. For both systems, we measured the total time needed to authorize a sensor operation as the time from the user input event to the authorization of the resulting operation request, corresponding to the last node of the delegation graph for \system.  We repeated the measurement 100 times for each system by varying the number of handoff events from 1 to 10. Each set of measurements was preceded by a priming run to remove any first-run effects. We then took an average of the middle 8 out of 10 such runs for each number of handoff events.
Figure~\ref{fig:benchmarks} shows that the overhead introduced by \system\ for the delegation graph enforcement is negligible, with the highest overhead observed being below 0.02\%. Thus, the slowdown is likely not to be noticeable by users. Indeed, none of our user study participants raised any concerns about discernible performance degradation or system slowdown.

\textbf{Ambiguity Prevention} -- Our fourth micro-benchmark was designed to measure the performance implications, in terms of delayed events, due to the ambiguity prevention mechanism. For this micro-benchmark, we selected the system UI (User Interface) process, which is one of the processes receiving the highest number of user input events, and the media server process, which receives the highest number of handoff events and therefore accesses system sensors with higher frequency than any other process. The time window for the construction of each delegation path was set to 150 ms. We generated 15,000 user input events with gaps randomly selected in the range [140-1,500]\footnote{\scriptsize Notice that to stress test our system, we selected a lower bound that is considerably lower than the maximum speed at which a user can possibly keep tapping on the screen ($\sim$210 ms).} ms. The time window and the gaps were selected based on data reported in Section \ref{subsec:field-based-study}. The generated input events caused 2,037 handoff events and 5,252 operation requests targeting system sensors (22,289 total scheduled events). The collected results indicated a total of 256 delayed events (1.15\% of the total events), with a maximum recorded delay of 9 ms.
Thus, the performance overhead introduced is negligible.

\textbf{Two-Level Queue Scheduling} -- Our fifth micro-benchmark was designed to measure the performance implications, in terms of additional delay for event delivery, due to the use of a two-level queue event scheduling. For this micro-benchmark, we compared two Android Nougat builds integrating \system\ with the ambiguity prevention mechanism enabled, but one of the two builds had the two-level queue scheduling disabled.  We used the Android UI/Application Exerciser Monkey to generate the same sequence of events for the same set of programs to guarantee fairness in the comparison between the two builds. For this micro-benchmark, we used the same settings as in our fourth micro-benchmark, but we varied the number of running applications, starting from 10 up to 100 apps\footnote{\scriptsize We stress tested our system. 21 is the average number of apps users install on personal smartphones, according to a statista.com review.} with an increment of 10 apps for each new run, for a total of 10 runs. The collected results reported an average of 311 delayed events (1.35\% of the total events) over the 10 runs, with a maximum recorded additional (compared to the build with the two-level queue scheduling disabled) delay of 5 ms, as summarized in Figure~\ref{fig:benchmarks}. Therefore, the two-level queue scheduling introduced  only a negligible additional delay. Also, we found that the maximum recorded delay slightly increased, at most 1 additional ms every 10 apps, \textit{for both builds} when the number of running apps increased. 

\textbf{Memory Requirement} -- We also recorded the average cache size required by \system\ to store both event mappings and authorized delegation graphs to be about 5.5 megabytes, for up to 1,000 programs.\footnote{\scriptsize Chosen among the most-downloaded Android apps from the Google Play Store and including all apps and system services shipped with the stock Android OS.} Therefore, \system\ required about 5.5 kilobytes of memory per program, which is a small amount of memory when compared to several gigabytes of storage available in modern systems. We ran the measurement 10 times and then took an average of the middle 8 out of 10 of such runs.

\section{Related Work}
\label{sec:related}

Researchers have studied security vulnerabilities related to Inter-Process Communications (IPCs), such as unauthorized use of intents, where adversaries can hijack activities and services by stealing intents, extensively \cite{chin2011analyzing, li2015iccta, felt2011permission, octeau2015composite, octeau2016combining, aafer2015hare}. In addition, several automated tools for IPC-related vulnerability analysis have been proposed. \textit{ComDroid} is a tool that parses the disassembled applications' code to analyze intent creation and transition for the identification of unauthorized intent reception and intent spoofing \cite{chin2011analyzing}. \textit{Efficient and Precise ICC discovery} (EPICC) is a more comprehensive static analysis technique for Inter-Component Communication (ICC)\footnote{\scriptsize Equivalent of IPCs for Android OS.} calls \cite{octeau2013effective}. It can identify ICC vulnerabilities due to intents that may be intercepted by malicious programs, or scenarios where programs expose components that can be launched via malicious intents. \textit{Secure Application INTeraction} (\textit{Saint}) \cite{ongtang2012semantically} extends the existing Android security architecture with policies that would allow programs to have more control to whom permissions for accessing their interfaces are granted and used at runtime. \textit{Quire} provides context in the form of provenance to programs communicating via Inter-Procedure Calls (IPC) \cite{dietz2011quire}. It annotates IPCs occurring within a system, so that the recipient of an IPC request can observe the full call sequence associated with it, before committing to any security-relevant decision.

Although effort has been made to analyze and prevent IPC-related vulnerabilities, none of the proposed approaches above tackled the problem from our perspective, i.e., instead of giving control to application developers, we must give control to users who are the real target for privacy violations by malicious programs. Toward our perspective, two defense mechanisms have been proposed by researchers. \textit{User-Driven Access Control} \cite{udac,udac2} proposes the use of access control gadgets, predefined by the operating systems and embedded into applications' code, to limit what operation can be associated with a specific user input event. \textit{AWare} \cite{petracca2017aware, Petracca2016AwareCA}, instead, proposes to bind each operation request, targeting sensitive sensors, to a user input event and obtains explicit authorization for the combination of operation request, user input event, and the user interface configuration used to elicit the event. Unfortunately, none of these mechanisms model interactions among programs via IPC. They only control how the user input event is consumed by the program receiving the user input event, therefore, they are not able to model input event delegation necessary to prevent the attack vectors discussed in this paper. Also, differently from prior work on permission re-delegation \cite{felt2011permission}, we do not rely on an over-restrictive defense mechanism that totally forbids permission re-delegation. This mechanism would block necessary interactions between programs even when the interaction is benign and expected by the user. Lastly, the use of decentralized information flow control policies, specified by developers and users \cite{nadkarni2016practical, nadkarni2013preventing}, requires the use of default-allow policies to ensure compatibility with legacy applications. However, a default-allow policy is unable to provide the same level of security provided by \system, which instead enforces a default-deny policy. Also, \system\ does not cause critical compatibility issues and does not require developers and users to label sensitive data or manually define per-process policies.

\section{Conclusion}
\label{sec:conclusion}

While a collaborative model allows the creation of useful, rich, and creative applications, it also introduces new attack vectors that can be exploited by adversaries. We have presented three classes of possible attack vectors exploitable by malicious programs, and proposed the \system\ authorization system to help mitigate them.
\system\ demonstrates that it is possible to prevent programs from abusing the collaborative model -- in the attempt to perform confused deputy, Trojan horse, or man-in-the-middle attacks -- by binding together, input event, handoff events, and sensor operation requests made by programs, and by requiring an explicit user authorization for the constructed delegation path.

\end{document}